\documentclass[12pt,letterpaper]{article}
\usepackage[margin=1.in]{geometry}
\usepackage[utf8]{inputenc}
\usepackage{amssymb}
\usepackage{amsmath}
\usepackage{physics}
\usepackage{graphicx}
\usepackage{url}
\usepackage{authblk}
\usepackage{xcolor}
\usepackage{enumitem}
\usepackage[most]{tcolorbox}
\PassOptionsToPackage{hyphens}{url}\usepackage[linktocpage=true, pdfencoding=auto, psdextra]{hyperref}

\usepackage[capitalize]{cleveref}
\allowdisplaybreaks
\setlist[itemize]{left=0pt..1em}

\usepackage[backend=biber,style=alphabetic,sorting=anyt]{biblatex}
\hypersetup{
	colorlinks=false,		
	citecolor=blue,		
	filecolor=black,		
	linkcolor=red,		
	urlcolor=black,		
	linkbordercolor =red, 	
	citebordercolor = blue,	
	urlbordercolor= blue	
}

\DeclareTotalTCBox{\Qtag}{ O{red} v !O{} }
{ fontupper=\ttfamily,nobeforeafter,tcbox raise base,arc=0pt,outer arc=0pt,top=0pt,bottom=0pt,left=0mm,right=0mm,leftrule=0pt,rightrule=0pt,toprule=0.3mm,bottomrule=0.3mm,boxsep=0.5mm,colback=#1!10!white,colframe=#1!50!black,#3}{#2}

\DeclareTotalTCBox{\Ttag}{ O{red} v !O{} }
{ fontupper=\ttfamily,nobeforeafter,tcbox raise base,top=0pt,bottom=0pt,left=0mm,right=0mm,leftrule=0pt,rightrule=0pt,toprule=0pt,bottomrule=0pt,boxsep=0.7mm,colback=#1!5!white,colframe=#1!50!black,#3}{#2}

\title{A Living Review of \\ Quantum Computing for Plasma Physics}

\author[1]{Óscar Amaro\thanks{\href{mailto:oscar.amaro@tecnico.ulisboa.pt}{oscar.amaro@tecnico.ulisboa.pt}}}
\author[2,3]{Diogo Cruz\thanks{\href{mailto:diogo.cruz@lx.it.pt}{diogo.cruz@lx.it.pt}}}
\affil{GoLP/Instituto de Plasmas e Fusão Nuclear, Instituto Superior Técnico, Lisboa, Portugal}
\affil{Instituto Superior Técnico, Universidade de Lisboa, Portugal}
\affil{Instituto de Telecomunicações, Portugal}

\date{\today}

\begin{document}

\maketitle

\begin{abstract}

Quantum Computing promises accelerated simulation of certain classes of problems, in particular in plasma physics. Given the nascent interest in applying quantum computing techniques to study plasma systems, a compendium of the relevant literature would be most useful. As a novel field, new results are common, and it is important for researchers to stay up-to-date on the latest developments. With this in mind, the goal of this document is to provide a regularly up-to-date and thorough list of citations for those developing and applying these quantum computing approaches to experimental or theoretical work in plasma physics. As a living document, it will be updated as often as possible to incorporate the latest developments. References are grouped by topic, both in itemized format and through the use of tags. We provide instructions on how to participate, and suggestions are welcome.
\end{abstract}

\section{Introduction}

Quantum Computing (QC) is a branch of computing that leverages on a set of resources provided by quantum mechanics that are classically unavailable: superposition, interference and entanglement. In some (but not all) situations, it is possible to construct a quantum algorithm (and a corresponding quantum gate circuit) that has a better complexity scaling than its classical counterpart. 

Consequently, interest in quantum computing has recently grown in several areas of engineering, physics, and other areas of research. Although still in its early stages, the growing body of research papers at the intersection of quantum computing and plasma physics makes it difficult to keep up to date with all of the latest developments. This is a challenge both for new researchers that plan to join the field and also for people who have worked in it for several years to put their contribution into the context of previous results.

To help address this issue, we have compiled a review of QC for plasma physics to provide a nearly comprehensive list of citations for papers that develop and apply QC (and derivations thereof) to problems in Plasma Physics, including theory and simulation/experiment. We have chosen the format of a \textit{living} review, which implies that it will be frequently updated and is open for community contributions. In this way, the review will continue to remain relevant as the field develops. It is worth noting that this living review does not qualify as a review of QC in general. We encourage the reader to look elsewhere in the literature for original research and reviews in areas of pure and applied quantum computing less relevant to plasma physics. This project was heavily inspired by a similar effort in the area of Machine Learning for High Energy Physics \cite{HEPML}.

The \textit{Living Review} (\url{https://qppqlivingreview.github.io/review}) also includes a list of standard (static) reviews within.
The references have been grouped into several categories to simplify the search process for the user (see \cref{sec:categories}). Some papers may be listed under multiple categories. The inclusion of a paper in the review does not indicate endorsement or validation of its content, as this is to be determined by the community and through peer review. The classification system may have limitations and we welcome feedback from the community on any papers that should be included, papers that have been misclassified or errors in citations or journal information.

The Living Review is hosted as a \href{https://github.com/QPPQLivingReview/review}{Github project}. It can be consulted either (see \cref{fig:descriptions}) as a \href{https://qppqlivingreview.github.io/review/}{Markdown/HTML webpage}, where the URL links for each reference are available; or as a \href{https://qppqlivingreview.github.io/review/review/review.pdf}{\LaTeX{} PDF file}, which is also downloadable directly from the webpage. The GitHub repository naturally also includes the BibTeX (.bib) file, which authors can freely use when writing new articles.

This pre-print will serve as a static reference to the review. Please check back before you upload your manuscript to arXiv to ensure that you include the latest work on the subject.

This paper is organized as follows: we briefly introduce the structure of the Living Review (\cref{sec:categories}); then we describe how one can contribute (\cref{sec:contribute}); finally we showcase a first version of the living review (\cref{sec:showcase}). The conclusions are in \cref{sec:concl}.

\section{Categories}
\label{sec:categories}

One of the main goals of this living review is to provide an easily searchable collection of references. These are presented in an itemized format, according to the type of problem the reference covers. In general, brief descriptions for each category of problem are provided. A single reference can be in more than one category, if applicable. 

\begin{figure}[!htpb]
 \centering
 \includegraphics[width=0.99\linewidth]{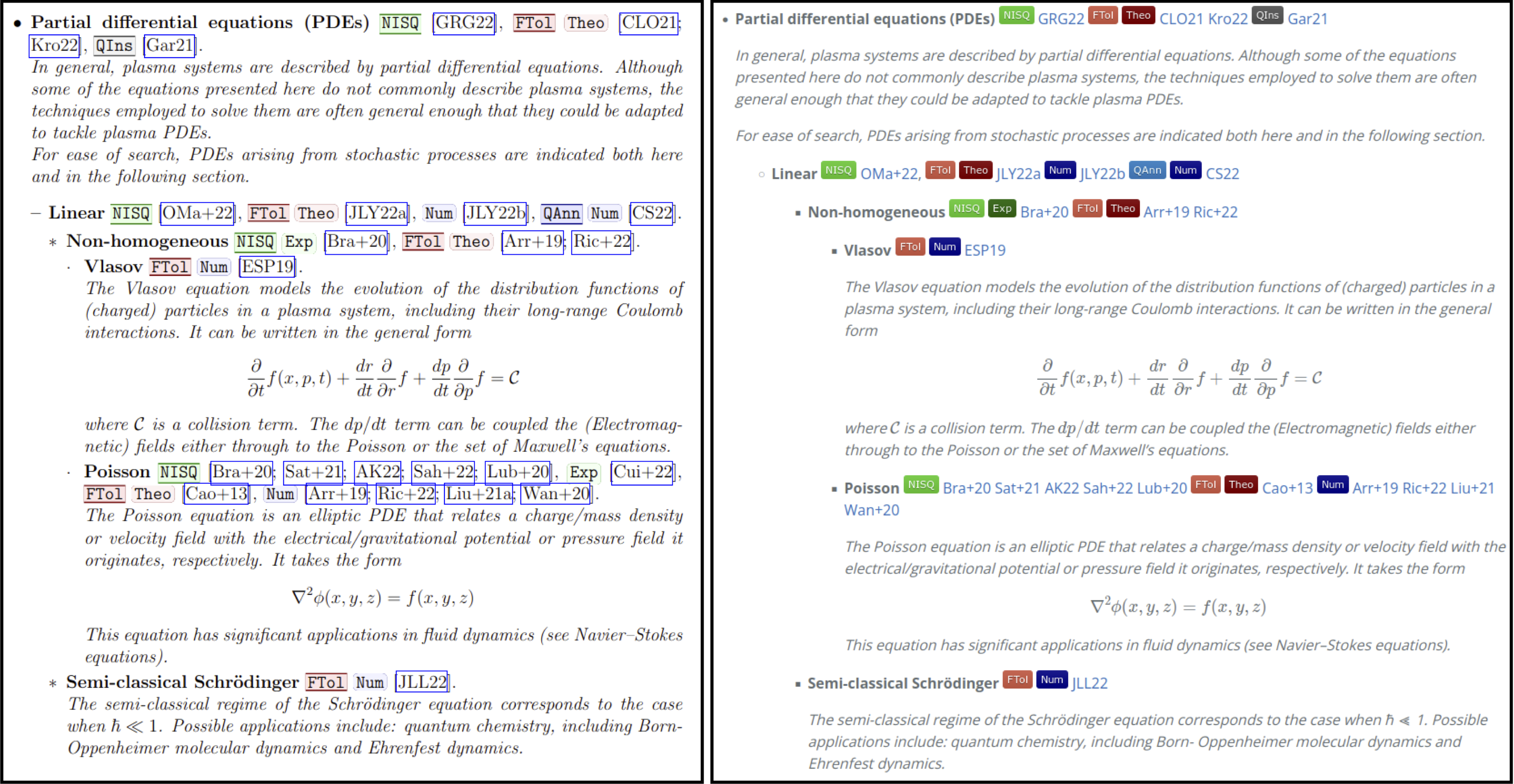}
 \caption{Snapshot of the PDF (left) and Markdown/HTML (right) form of the review with hyperlinks to the papers.}
 \label{fig:descriptions}
\end{figure}

For the same type of problem, it is not uncommon for each researcher to be mostly interested in specific types of quantum computing techniques, applicable in specific contexts. Therefore, to facilitate search, we propose the use of tags to mark references with the context under which the problem was considered, as shown in \cref{fig:descriptions}.

The following categories are used to organize the papers:
\begin{itemize}
    \item \Qtag[green]{NISQ} noisy-intermediate scale quantum computing;
    \item \Qtag[red]{FTol} fault-tolerant quantum computing;
    \item \Qtag[blue]{QAnn} quantum annealing;
    \item \Qtag[gray]{QIns} quantum-inspired.
 \end{itemize}

 Moreover, we also use secondary tags to indicate the type of analysis performed:
\begin{itemize}
    \item \Ttag{Theo} is a theoretical tag solely for the papers with analytical results, and no considerable numerical or experimental results;
    \item  \Ttag[blue]{Num} marks papers with numerical simulations, but no experimental results run on quantum devices;
    \item \Ttag[green]{Exp} marks papers with displayed experimental results.
\end{itemize}
We may omit these tags if the paper is referenced and tagged in a subsequent subsection, or if none of them are clearly applicable.

Please note that this choice of structuring is somewhat arbitrary and may change as the field evolves.

\begin{figure}[!htpb]
 \centering
 \includegraphics[width=0.99\linewidth]{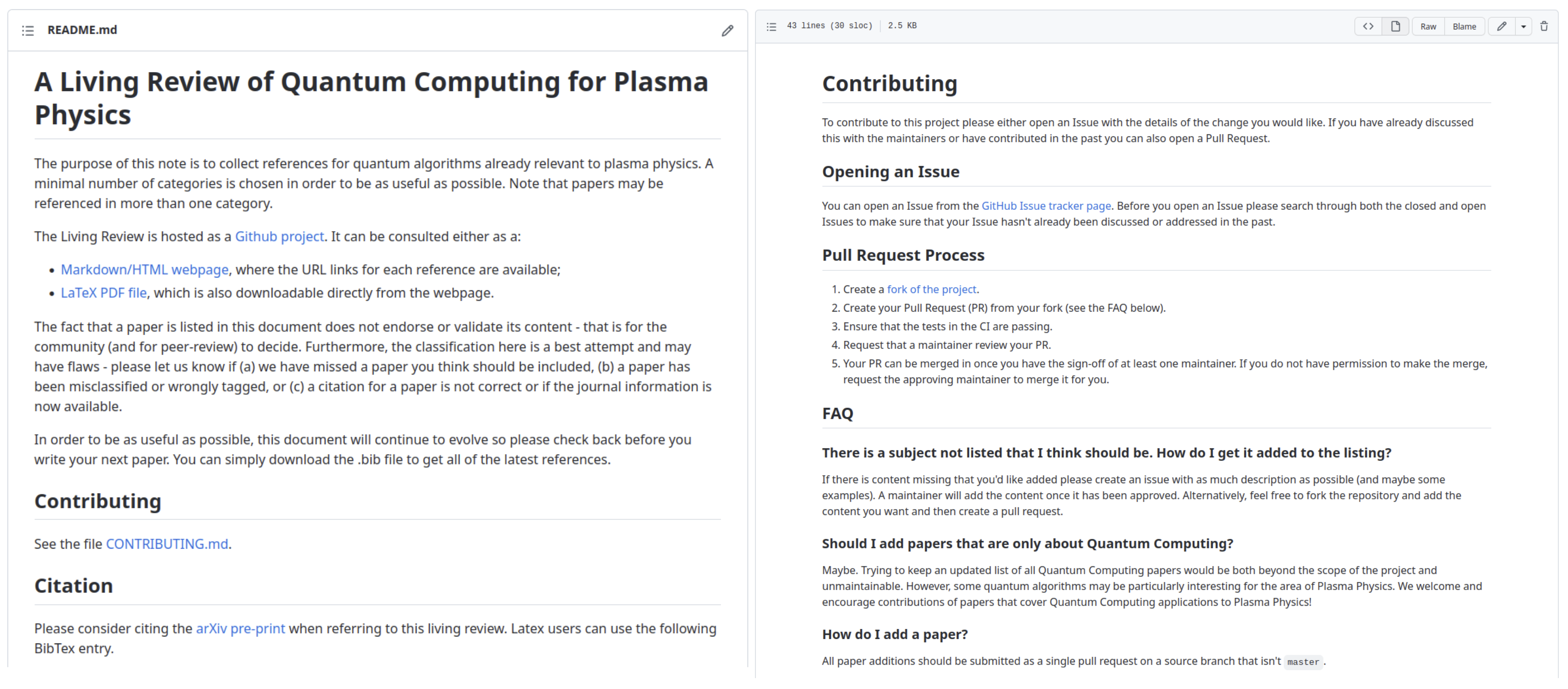}
 \caption{A printscreen of the \texttt{README.md} (left) and \texttt{CONTRIBUTING.md} (right) document describing the repository's structure and detailing the guidelines for contributions from non-maintainers to the review.}
 \label{fig:CONTRIBUTING}
\end{figure}

\section{Contributing}
\label{sec:contribute}

The community can also become involved in the process of reviewing and expanding this living document. The main channel for contributions is through a pull request (PR) to the review's \href{https://github.com/QPPQLivingReview/review}{GitHub project}. A frequently asked questions (FAQ) and pull-request guide can be found in the repository, with detailed instructions on the recommended procedures and workflow.

To help steer new contributions and ensure a smooth PR process and review with the maintainers, a contributions guide is located in the project's Git repository in the form of a \texttt{CONTRIBUTING.md} document (see \cref{fig:CONTRIBUTING}) --- a project staple in the Open Source community.

Furthermore, suggestions for new features can be submitted to the Living Review by creating a \href{https://github.com/QPPQLivingReview/review/issues}{GitHub issue} as documented in the project's \texttt{CONTRIBUTING.md}.

\section{Showcase of the living review}\label{sec:showcase}

For reference, in this section we showcase the starting version of the Living Review, as it stood at the time it was first made public. As previously indicated, a continuously updated version of it may be found in its associated \href{https://github.com/QPPQLivingReview/review}{GitHub project}.

\begin{itemize} 

	\item \textbf{Modern Reviews}
		
    Below are links to (static) general and specialized reviews.
    \begin{itemize} 
        \item Applications of Quantum Computing to Plasma Simulations \cite{dodinApplicationsQuantumComputing2021}.
        \item Quantum Computing for Fusion Energy Science Applications \cite{josephQuantumComputingFusion2022a}.
    \end{itemize}
    
	\item \textbf{System of linear equations} \Qtag[green]{NISQ} \Ttag[blue]{Num}~\cite{huangNeartermQuantumAlgorithms2021} \Ttag[green]{Exp}\cite{bravo-prietoVariationalQuantumLinear2020,xuVariationalAlgorithmsLinear2021}, \Qtag[red]{FTol} \Ttag[red]{Theo}~\cite{harrowQuantumAlgorithmLinear2009,claderPreconditionedQuantumLinear2013,childsQuantumAlgorithmSystems2017,wangEfficientQuantumAlgorithms2022}, \Qtag[blue]{QAnn} \Ttag{Num}~\cite{subasiQuantumAlgorithmsSystems2019}, \Ttag[green]{Exp}~\cite{borleHowViableQuantum2022}, \Qtag[gray]{QIns} \Ttag[red]{Theo}~\cite{shaoFasterQuantuminspiredAlgorithms2021}.
    
    Many problems in plasma physics may be formulated, either exactly or approximately, as a problem of the form $A x = b$, where $A$ and $b$ encode the information about the system (including its initial conditions, if applicable), and the goal is to compute $x$, which encodes the desired data.
 
	\item \textbf{System of nonlinear equations} ~\Qtag[red]{FTol} \Ttag{Theo}~\cite{dodinQuantumComputationNonlinear2021} \Ttag[blue]{Num}~\cite{xueQuantumNewtonMethod2021,xueQuantumAlgorithmSolving2022}.
    
    Nonlinear equations depend nonlinearly on $x$, and they are generally much harder to solve. As quantum mechanics is inherently linear, many techniques rely on mapping the original nonlinear problem to a (usually approximate) linear one, that is easier to solve.
    \begin{itemize} 
        \item \textbf{System of polynomial equations} \Qtag[blue]{QAnn} \Ttag[green]{Exp}~\cite{changQuantumAnnealingSystems2019}.

        System where each equation is a polynomial.
    \end{itemize}
  
	\item \textbf{Ordinary differential equations (ODEs)} \Qtag[blue]{QAnn} \Ttag[green]{Exp}~\cite{zangerQuantumAlgorithmsSolving2021}.
    
    Many plasma systems can be described by ordinary differential equations. Some techniques attempt to solve them directly, while others map the ODE to (larger) systems of linear equations, and solve those.
    \begin{itemize} 
        \item \textbf{Linear} \Qtag[red]{FTol} \Ttag[red]{Theo}~\cite{berryHighorderQuantumAlgorithm2014,berryQuantumAlgorithmLinear2017,childsQuantumSpectralMethods2020,fangTimemarchingBasedQuantum2022,jinQuantumSimulationPartial2022}, \Ttag[blue]{Num}~\cite{jinQuantumSimulationPartial2022a}, \Qtag[blue]{QAnn} \Ttag[green]{Exp}~\cite{zangerQuantumAlgorithmsSolving2021}.

        As quantum mechanics is inherently linear, linear ODEs are often more straightforward to solve with quantum computers than nonlinear ones.
        
        \begin{itemize} 
            \item \textbf{Second-order} \Qtag[blue]{QAnn} \Ttag[green]{Exp}~\cite{srivastavaBoxAlgorithmSolution2019}.

            The highest derivative appearing in these ODEs is the second derivative.
            
            \begin{itemize} 
                \item \textbf{Quantum harmonic oscillator} \Qtag[red]{FTol} \Ttag[blue]{Num}~\cite{ricardoAlternativesNonhomogeneousPartial2022}.

                Here, the time-independent Schrödinger equation has a Hamiltonian with a potential proportional to $x^2$.

                \item \textbf{Laguerre} \Qtag[blue]{QAnn} \Ttag[blue]{Num}~\cite{criadoQadeSolvingDifferential2022}.

                The Laguerre equation is of the form $$ x \dfrac{\mathrm{d}^2 y}{\mathrm{d} x^2} + (1-x) \dfrac{\mathrm{d} y}{\mathrm{d} x} + n ~y = 0$$
            \end{itemize}
        \end{itemize}
        \item \textbf{Nonlinear} ~\Qtag[green]{NISQ} ~\cite{kyriienkoSolvingNonlinearDifferential2021, shiSimulatingNonnativeCubic2021} ~\Qtag[red]{FTol} \Ttag{Theo}~\cite{leytonQuantumAlgorithmSolve2008,dodinQuantumComputationNonlinear2021,lloydQuantumAlgorithmNonlinear2020} \Ttag[blue]{Num}~\cite{jinQuantumSimulationPartial2022a,liuEfficientQuantumAlgorithm2021,suranaCarlemanLinearizationBased2022b}, ~\Qtag[blue]{QAnn} \Ttag[blue]{Num}~\cite{zangerQuantumAlgorithmsSolving2021}.

        There is no general reliable procedure to solve nonlinear ODEs, but some methods have been proposed.
    \end{itemize}
  
	\item \textbf{Partial differential equations (PDEs)} \Qtag[green]{NISQ}~\cite{garcia-molinaQuantumFourierAnalysis2022}, \Qtag[red]{FTol} \Ttag[red]{Theo}~\cite{childsHighprecisionQuantumAlgorithms2021,kroviImprovedQuantumAlgorithms2022}, \Qtag[gray]{QIns}~\cite{garcia-ripollQuantuminspiredAlgorithmsMultivariate2021}.
 
    In general, plasma systems are described by partial differential equations. Although some of the equations presented here do not commonly describe plasma systems, the techniques employed to solve them are often general enough that they could be adapted to tackle plasma PDEs.
    
    For ease of search, PDEs arising from stochastic processes are indicated both here and in the following section.
		\begin{itemize} 
			\item \textbf{Linear} \Qtag[green]{NISQ}~\cite{omalleyNeartermQuantumAlgorithm2022}, \Qtag{FTol} \Ttag{Theo}~\cite{jinQuantumSimulationPartial2022}, \Ttag[blue]{Num}~\cite{jinQuantumSimulationPartial2022a}, \Qtag[blue]{QAnn} \Ttag[blue]{Num}~\cite{criadoQadeSolvingDifferential2022}.
				\begin{itemize} 
					\item \textbf{Non-homogeneous} \Qtag[green]{NISQ} \Ttag[green]{Exp}~\cite{bravo-prietoVariationalQuantumLinear2020}, \Qtag[red]{FTol} \Ttag[red]{Theo}~\cite{arrazolaQuantumAlgorithmNonhomogeneous2019,ricardoAlternativesNonhomogeneousPartial2022}.
						\begin{itemize} 
                            \item \textbf{Vlasov} \Qtag[red]{FTol}~\Ttag[blue]{Num}~\cite{engelQuantumAlgorithmVlasov2019}.

                            The Vlasov equation models the evolution of the distribution functions of (charged) particles in a plasma system, including their long-range Coulomb interactions. It can be written in the general form
                            $$ \dfrac{\partial }{\partial t} f(x,p,t) + \dfrac{d r}{d t} \dfrac{\partial }{\partial r} f + \dfrac{d p}{d t} \dfrac{\partial }{\partial p} f = \mathcal{C}$$
                            where $\mathcal{C}$ is a collision term. The $dp/dt$ term can be coupled the (Electromagnetic) fields either through to the Poisson or the set of Maxwell's equations.
                            
							\item \textbf{Poisson} \Qtag[green]{NISQ}~\cite{bravo-prietoVariationalQuantumLinear2020,satoVariationalQuantumAlgorithm2021,aliPerformanceStudyVariational2022,sahaAdvancingAlgorithmScale2022,lubaschVariationalQuantumAlgorithms2020}, \Ttag[green]{Exp}~\cite{cuiOptimizationNoiseAnalysis2022}, \Qtag[red]{FTol} \Ttag[red]{Theo}~\cite{caoQuantumAlgorithmCircuit2013}, \Ttag[blue]{Num}~\cite{arrazolaQuantumAlgorithmNonhomogeneous2019,ricardoAlternativesNonhomogeneousPartial2022,liuVariationalQuantumAlgorithm2021,wangQuantumFastPoisson2020}.

                          The Poisson equation is an elliptic PDE that relates a charge/mass density or velocity field with the electrical/gravitational potential or pressure field it originates, respectively. It takes the form
                        $$ \nabla^2 \phi(x,y,z) = f(x,y,z) $$
                        This equation has significant applications in fluid dynamics (see Navier–Stokes equations).
       
						\end{itemize}

                    \item \textbf{Semi-classical Schrödinger} \Qtag{FTol} \Ttag[blue]{Num}~\cite{jinQuantumSimulationSemiclassical2022}.
                    
                    The semi-classical regime of the Schrödinger equation corresponds to the case when $\hbar \ll 1$. Possible applications include: quantum chemistry, including Born-Oppenheimer molecular dynamics and Ehrenfest dynamics.
                    
                    \item \textbf{Time-dependent Schrödinger} \Qtag[green]{NISQ}~\cite{joubert-doriolVariationalApproachLinearly2022} \Qtag[red]{FTol} \Ttag[blue]{Num}~\cite{jinQuantumSimulationSemiclassical2022}.

                    Many problems can be mapped to the Schrödinger equation $$\mathrm{i}\hbar \dv{t} \ket{\psi(t)} = H\ket{\psi(t)}$$ described by some Hamiltonian $H$. Solving the Schrödinger equation can often be done much more efficiently with quantum computers.

                    \item \textbf{Parabolic} \Qtag[gray]{QIns}~ \Ttag[blue]{Num} \cite{patelQuantumInspiredTensorNeural2022}.
                    \begin{itemize} 

                        \item \textbf{Heat/Convection} \Qtag[green]{NISQ}~\cite{leongVariationalQuantumEvolution2022,albinoSolvingPartialDifferential2022}, \Ttag[blue]{Num}~\cite{liuApplicationVariationalHybrid2022}, \Qtag[red]{FTol} \Ttag[red]{Theo}~\cite{lindenQuantumVsClassical2022,jinQuantumSimulationPartial2022,jinTimeComplexityAnalysis2022a} \Ttag[blue]{Num}~\cite{jinQuantumSimulationPartial2022a}.

                        The prototypical parabolic linear PDE. This equation describes the diffusion of heat in a material

                        $$\partial_t u(t,x) = \alpha \Delta u $$
                        where $\alpha$ is the thermal diffusivity. Its applications are of fundamental importance in most branches of physics and engineering.
                    
            			\item \textbf{Black-Scholes} \Qtag[green]{NISQ}~\cite{fontanelaShortCommunicationQuantum2021,miyamotoPricingMultiAssetDerivatives2022}, \Qtag[red]{FTol} \Ttag[blue]{Num}~\cite{jinQuantumSimulationPartial2022a,anQuantumacceleratedMultilevelMonte2021}.
               
                        The Black–Scholes equation is a PDE which describes the price of the option $V(S,t)$ over time $t$ and price of underlying asset $S(t)$,$r$ is the "force of interest", $\mu$ is the annualized drift rate of $S$, and $\sigma$ is the standard deviation of the stock's returns.
                        $$
                        \frac{\partial V}{\partial t}+\frac{1}{2} \sigma^2 S^2 \frac{\partial^2 V}{\partial S^2}+r S \frac{\partial V}{\partial S}-r V=0
                        $$

                        \item \textbf{Fokker-Planck} \Qtag[red]{FTol} \Ttag[blue]{Num}~\cite{jinQuantumSimulationPartial2022a}, \Qtag[gray]{QIns} \Ttag[blue]{Num}~\cite{garcia-ripollQuantuminspiredAlgorithmsMultivariate2021}.
                        
                        The Fokker-Planck equation for the probability density $p(x,t)$ can be written as
                        $$ \dfrac{\partial }{\partial t} p(x,t)  = - \dfrac{\partial }{\partial x} \left( \mu(x,t)~p(x,t)  \right) + \dfrac{\partial^2 }{\partial x^2} \left( D(x,t)~p(x,t)  \right)$$
                        where $\mu$ and $D$ are the drift and diffusion coefficients (which may be time-dependent).
                        
                        \item \textbf{Hamilton-Jacobi-Bellman} \Qtag[gray]{QIns} \Ttag[blue]{Num}\cite{patelQuantumInspiredTensorNeural2022}.
                    \end{itemize}

                    \item \textbf{Hyperbolic/Wave-related} \Qtag{FTol} \Ttag{Theo}~\cite{jinTimeComplexityAnalysis2022a,jinQuantumAlgorithmsComputing2022}.
                         \begin{itemize} 
                             \item \textbf{Wave} \Qtag{FTol} \Ttag[blue]{Num}~\cite{costaQuantumAlgorithmSimulating2019,suauPracticalQuantumComputing2021}, \Qtag[blue]{QAnn} \Ttag[blue]{Num}~\cite{criadoQadeSolvingDifferential2022}.

                             The prototypical hyperbolic equation in physics, it describes oscillatory and propagating perturbations in a medium

                             $$ \partial_{tt} u(t,x) = c^2 \nabla^2 u$$
                         
                            \item \textbf{Maxwell's} \Qtag{FTol} \Ttag{Theo}~\cite{costaQuantumAlgorithmSimulating2019}, \Ttag[blue]{Num}~\cite{novikauQuantumSignalProcessing2022,novikauSimulationLinearNonHermitian2022}.
                            
                            The Maxwell equations are a set of coupled PDEs that are foundational to the modeling of electromagnetic phenomena, which important applications in fundamental physics, classical optics and electric circuits.
                            
                            $$\nabla \cdot \mathbf{E}=\frac{\rho}{\varepsilon_0}, \nabla \cdot \mathbf{B}=0, \nabla \times \mathbf{E}=-\frac{\partial \mathbf{B}}{\partial t}, \nabla \times \mathbf{B}=\mu_0\left(\mathbf{J}+\varepsilon_0 \frac{\partial \mathbf{E}}{\partial t}\right)$$

                           \item \textbf{Klein-Gordon} \Qtag{FTol} \Ttag{Theo}~\cite{costaQuantumAlgorithmSimulating2019}.

                            The Klein-Gordon relativistic wave equation is a 2nd order equation both in space and time

                            $$
                                \left(\frac{1}{c^2} \frac{\partial^2}{\partial t^2}-\nabla^2+\frac{m^2 c^2}{\hbar^2}\right) \psi(t, \mathbf{x})=0
                            $$
                           
                           \item \textbf{Helmholtz} \Qtag[green]{NISQ} \Ttag[blue]{Num} ~\cite{eweVariationalQuantumBasedSimulation2022}.

                            The Helmholtz equation is the eigenvalue problem for the Laplace operator
                            
                           $$ \nabla^2 f = - k^2 f$$
                           
                           Important applications include wave-like phenomena and diffusion processes.
                           
                         \end{itemize}  

				\end{itemize}

            \item \textbf{Nonlinear} \Qtag{FTol} \Ttag{Theo}~\cite{jinQuantumAlgorithmsComputing2022,linKoopmanNeumannMechanics2022}.
            
            \begin{itemize} 
            
    			    \item \textbf{Evolution equation} \Qtag[green]{NISQ}~ \Ttag[red]{Theo}~\cite{leongVariationalQuantumEvolution2022}.
                
                    General class of PDEs with time-domain.
            
					\item \textbf{Vlasov-Poisson} \Qtag[gray]{QIns} \Ttag[blue]{Num}~\cite{yeQuantuminspiredMethodSolving2022}. 

                    The coupled Vlasov-Poisson system describes the self-consistent evolution of charges and their (electrical) potentials. Because of this, the system is nonlinear.
                    
                    \item \textbf{Schrödinger-Poisson} \Qtag[green]{NISQ}~\Ttag[blue]{Num}~\cite{moczCosmologicalSimulationsDark2021}.

                    Similar to the Vlasov-Poisson system, the Schrödinger-Poisson equation attempts to describe the self-consistent evolution of a wave-function and a potential.

                    \item \textbf{Nonlinear-Schrödinger} \Qtag[green]{NISQ} \Ttag[blue]{Num}~\cite{garcia-molinaQuantumFourierAnalysis2022}, \Ttag[green]{Exp}~\cite{lubaschVariationalQuantumAlgorithms2020}.
                    
                    Similar to the Schrödinger-Poisson system, however the potential is created by the square absolute value of the wave-function $V=|\psi|^2$, which induces a cubic nonlinearity in the equations.
                    
                   \item \textbf{Burger's} \Qtag[green]{NISQ} \Ttag[green]{Exp}~\cite{zylbermanHybridQuantumClassicalAlgorithm2022}, \Qtag{FTol} \Ttag[blue]{Num}~\cite{ozSolvingBurgersEquation2021}.
                   
                   This nonlinear PDE with a quadratic nonlinearity is of fundamental importance to fluid dynamics and in particular in plasma physics. It is also a prototypical equation in turbulence related studies, and can be written in the form

                   $$\dfrac{\partial }{\partial t} u(x,t) +u \dfrac{\partial u}{\partial x} = \nu \dfrac{\partial^2 u}{\partial x^2}$$
       
    			\item \textbf{Reaction-diffusion} \Qtag[green]{NISQ}~ \Ttag[blue]{Num}~\cite{leongVariationalQuantumEvolution2022} \Qtag[green]{NISQ}~ \Ttag[green]{Exp}~\cite{demirdjianVariationalQuantumSolutions2022}, \Qtag{FTol} \Ttag[blue]{Num}~\cite{anEfficientQuantumAlgorithm2022}.
       
                Reaction–diffusion equations describe transport/diffusion of substances and their transformation into other substances, for example to model the concentration of chemical components.
                
                $$ \partial_t p = D \nabla^2 q + R(q)$$
                where $D$ is the diffusion coefficient and $R$ is the local reaction rate.      
    
                \item \textbf{Navier-Stokes} \Qtag{FTol} \Ttag[blue]{Num}~\cite{gaitanFindingFlowsNavier2020}.
                
                The Navier-Stokes equations describe the motion of viscous fluids. They describe conservation of mass and momentum and often require equations of state for pressure, temperature and density to close the system of equations.

                $$ \partial_t \rho (x,t) + \nabla \cdot (\rho u) = 0, \frac{\partial}{\partial t}(\rho \mathbf{u})+\nabla \cdot(\rho \mathbf{u} \otimes \mathbf{u})=-\nabla p+\mu \nabla^2 \mathbf{u}+\frac{1}{3} \mu \nabla(\nabla \cdot \mathbf{u})+\rho \mathbf{g}$$
                
                \begin{itemize} 
                    \item \textbf{Incompressible} \Qtag[green]{NISQ}~\cite{leongVariationalQuantumEvolution2022}, \Qtag[gray]{QIns} \Ttag[blue]{Num}~\cite{lapworthHybridQuantumClassicalCFD2022}.

                    The incompressible Navier-Stokes equations can be applied for low enough Mach numbers, and cannot be used to accurately simulate density or pressure waves like sound or shock waves. The fluid density is considered constant $\rho = \rho_0$.
                \end{itemize}
    
                \item \textbf{Hamilton-Jacobi} \Qtag{FTol} \Ttag{Theo}~\cite{jinTimeComplexityAnalysis2022,jinQuantumAlgorithmsComputing2022}.
                
                A particular case of the Hamilton-Jacobi-Bellman. The Hamilton-Jacobi is an alternative formulation of classical mechanics, where given a Hamiltonian $H( q , p , t )$ of a mechanical system, the Hamilton–Jacobi equation is a first-order, non-linear PDE for the Hamilton's principal function $S$. One of its advantages is in efficiently identifying conserved quantities of mechanical systems.

                \item \textbf{Black-Scholes-Barenblatt} \Qtag[gray]{QIns} \Ttag[blue]{Num}\cite{patelQuantumInspiredTensorNeural2022}.
                
                The Black-Scholes-Barenblatt equation is a nonlinear extension to the Black-Scholes equation, which models uncertain volatility and interest rates derived from the Black-Scholes equation.

                \item \textbf{Stefan problems} \Qtag{FTol} \Ttag{Theo}~\cite{sarsengeldinHybridClassicalQuantumFramework2022}.
                
                Stefan problems are a particular kind of boundary value problems for a system of PDEs in which the boundary between the phases can move with time.
                
            \end{itemize}
		\end{itemize}

    \item \textbf{Stochastic processes} \Qtag[green]{NISQ} \Ttag[blue]{Num}~\cite{kuboVariationalQuantumSimulations2020, alghassiVariationalQuantumAlgorithm2022}, \Qtag{FTol} \Ttag[blue]{Num}~\cite{anQuantumacceleratedMultilevelMonte2021}.

    (Integro-)differential equations in which one or more of the terms is a stochastic process, leading to a solution which is stochastic in nature. Stochastic Differential Equations (SDEs) can be used to model physical systems subject to thermal fluctuations.\\
    Through the Feynman-Kac formula, many common SDEs can be reduced to solving a PDE for the probability density of interest, as is the case for the Fokker-Planck equation.\\
    For ease of search, PDEs arising from stochastic processes are indicated both here and in the previous section.

   \begin{itemize} 
        \item \textbf{Fokker-Planck} \Qtag[red]{FTol} \Ttag[blue]{Num}~\cite{jinQuantumSimulationPartial2022a}, \Qtag[gray]{QIns} \Ttag[blue]{Num}~\cite{garcia-ripollQuantuminspiredAlgorithmsMultivariate2021}.
        
        The Fokker-Planck equation for the probability density $p(x,t)$ can be written as
        $$ \dfrac{\partial }{\partial t} p(x,t)  = - \dfrac{\partial }{\partial x} \left( \mu(x,t)~p(x,t)  \right) + \dfrac{\partial^2 }{\partial x^2} \left( D(x,t)~p(x,t)  \right)$$
        where $\mu$ and $D$ are the drift and diffusion coefficients (which may be time-dependent).

        \item \textbf{Linear Boltzmann/Rate equation} \Qtag[red]{FTol} \Ttag[blue]{Num}~\cite{jinQuantumSimulationPartial2022a}.

        The linear Boltzmann or Rate equation  is a stochastic integro-differential equation  for the probability density $p(x,t)$ can be written as
        $$ \dfrac{d }{d t} p(x,t)  = \int p(x',t) W(x,x') dx' - p(x,t) \int W(x',x) dx' $$
        where $W(x,x')$ is the probability rate of transition from state $x'$ to state $x$. 
        where $dp/dt$ can include partial derivatives of $p(x,t)$.
        
   \end{itemize}

    \item \textbf{Other techniques}
    \begin{itemize} 
                
        \item \textbf{Linear embedding of nonlinear dynamical systems} \Qtag[red]{FTol} \Ttag[red]{Theo} \cite{engelLinearEmbeddingNonlinear2021,jinTimeComplexityAnalysis2022} \Ttag[blue]{Num} \cite{liuEfficientQuantumAlgorithm2021}.

        Several linear embedding of nonlinear dynamical systems have been developed to extend the class of problems that can be tackled by quantum computers. These include Koopman–von Neumann formulation, Quantum nonlinear Schrödinger linearization formulation and Carleman linearization, amongst others.
        
        \item \textbf{Koopman–von Neumann formulation} \Qtag[red]{FTol} \Ttag[red]{Theo} \cite{josephKoopmanNeumannApproach2020,jinTimeComplexityAnalysis2022}, \Ttag[blue]{Num}~\cite{linKoopmanNeumannMechanics2022}.

        Koopman–von Neumann mechanics is a description of classical mechanics embedded in a Hilbert space. The dynamical equation can be written as $\mathrm{i} \partial_t \psi=\mathcal{H}_{\mathrm{KvN}}~ \psi $, where the operator $\mathcal{H}_{\mathrm{KvN}}=-\mathrm{i} \sum_j\left(F_j \frac{\partial}{\partial x_j}+\frac{1}{2} \frac{\partial F_j}{\partial x_j}\right)$. Furthermore, the probability density is interpreted as  $\rho = |\psi|^2$. Applications may include the Vlasov-Maxwell coupled system of equations.
            
        \item \textbf{Quantum nonlinear Schrödinger linearization formulation} \Qtag[red]{FTol} \Ttag[red]{Theo} \cite{lloydQuantumAlgorithmNonlinear2020}.

        Formulation of ODEs/PDEs of the type $ dx/dt +f(x) x = b(t) $, with $f = x^{\dagger \otimes m} F x^{\otimes m}$, as nonlinear Schrödinger equations. Potential applications of the method may include the Navier-Stokes equation, plasma hydrodynamics, epidemiology.
    
        \item \textbf{Madelung transform for nonlinear relativistic fluids} \Ttag[blue]{Num}~\cite{hatifiQuantumWalkHydrodynamics2019}, \Qtag[green]{NISQ} \Ttag[green]{Exp}~\cite{zylbermanHybridQuantumClassicalAlgorithm2022}.

        The Madelung equations are an alternative formulation to the Schrödinger equation, written in terms of hydrodynamical variables, with the addition of the Bohm quantum potential $Q$
        
        $$\partial_t \rho_m+\nabla \cdot\left(\rho_m \mathbf{u}\right)=0, ~\frac{d \mathbf{u}}{d t}=\partial_t \mathbf{u}+\mathbf{u} \cdot \nabla \mathbf{u}=-\frac{1}{m} \nabla(Q+V)$$

        Applications include modeling of shocks in plasmas.
    
        \item \textbf{Finite element method} \Qtag{FTol} \Ttag{Theo}~\cite{montanaroQuantumAlgorithmsFinite2016}.
        
        This approach relies on dividing a system/domain into smaller regions called finite elements. The discretization process applied to a boundary value problem leads to a system of algebraic equations, which can be less computationally expensive to resolve than the original PDE. Important applications include fluid flow, heat transfer, and electromagnetic potentials.
    
        \item \textbf{Lattice Boltzmann algorithms} \Qtag[red]{Ftol}~\Ttag[blue]{Num}~\cite{ljubomirQuantumAlgorithmNavier2022}.

        Originally developed as a classical algorithm, this approach can be used to simulate fluid dynamics without having to solve the Navier–Stokes equations directly. The fluid density is represented on a lattice and evolves in time with streaming and collision processes. One of the advantages of the method is its efficiency/scalability in parallel architectures.
        
        \item \textbf{Quantum lattice algorithms} \Qtag[gray]{QIns} \Ttag[blue]{Num}~\cite{andersonCommentsUnitaryQubit2022,koukoutsisDysonMapsUnitary2022,oganesovEffectFourierTransform2018,ramReflectionTransmissionElectromagnetic2021,vahalaBuildingThreedimensionalQuantum2020,vahalaEffectPauliSpin2020,vahalaOneTwodimensionalQuantum2021,vahalaOneTwodimensionalQuantum2021a,vahalaQuantumLatticeRepresentation2022a,vahalaQubitUnitaryLattice2020,vahalaQubitUnitaryLattice2020a,vahalaTwoDimensionalElectromagnetic2021,vahalaUnitaryQuantumLattice2020,vahalaUnitaryQubitLattice2019,yepezEfficientAccurateQuantum2002,yepezRelativisticPathIntegral2005,yepezQuantumLatticeGas2016,vahalaQuantumLatticeGas2003,vahalaUnitaryQubitLattice2011,vahalaUnitaryQuantumLattice2010,oganesovBenchmarkingDiracgeneratedUnitary2016,oganesovImaginaryTimeIntegration2016,oganesovUnitaryQuantumLattice2015,shiSimulationsRelativisticQuantum2018}.
        \\ Highly parallelizable approach amenable to classical supercomputers, allowing the study of (Klein-Gordon-)Maxwell's equations, the Gross-Pitaevski equation, the nonlinear Schrödinger equation, and the KdV equation. In some cases, the method may also be suitable for fault-tolerant quantum computers.
    \end{itemize}

\end{itemize}

\section{Conclusions}
\label{sec:concl}

This paper presents the \emph{Living Review of Quantum Computing for Plasma Physics}, which will be regularly updated with the latest research papers from open repositories and scientific journals. The community is invited to participate in the project at any time. Quantum computing has the potential to significantly speed up simulations of plasma physics systems, and it is hoped that this Living Review will serve as a valuable resource for keeping up with the rapid advances in the field.

\section{Acknowledgments}

We would like to thank Marija Vranic, Akshat Kumar, Luís Oliveira e Silva and Yasser Omar for fruitful discussions and the support given. We also thank Miguel Murça, Duarte Magano, Sagar Pratapsi, Gonçalo Vaz, Pablo Bilbao, Chiara Badiali and Tiago Martins for useful feedback.
DC acknowledges the support from FCT through scholarship UI/BD/152301/2021 and OA acknowledges the support from FCT through scholarship UI/BD/153735/2022.

\printbibliography

@misc{HEPML,
    doi = {10.48550/ARXIV.2102.02770},
    url = {https://arxiv.org/abs/2102.02770},
    author = {Feickert, Matthew and Nachman, Benjamin},
    title = {A Living Review of Machine Learning for Particle Physics},
    publisher = {arXiv},
    year = {2021},
    copyright = {Creative Commons Attribution 4.0 International}
  }

@misc{albinoSolvingPartialDifferential2022,
  title = {Solving Partial Differential Equations on Near-Term Quantum Computers},
  author = {Albino, Anton Simen and Jardim, Lucas Correia and Knupp, Diego Campos and Neto, Antonio Jose Silva and Pires, Otto Menegasso and Nascimento, Erick Giovani Sperandio},
  year = {2022},
  month = aug,
  number = {arXiv:2208.05805},
  eprint = {2208.05805},
  eprinttype = {arxiv},
  primaryclass = {physics, physics:quant-ph},
  publisher = {{arXiv}},
  doi = {10.48550/arXiv.2208.05805},
  archiveprefix = {arXiv}
}

@article{alghassiVariationalQuantumAlgorithm2022,
  title = {A Variational Quantum Algorithm for the {{Feynman-Kac}} Formula},
  author = {Alghassi, Hedayat and Deshmukh, Amol and Ibrahim, Noelle and Robles, Nicolas and Woerner, Stefan and Zoufal, Christa},
  year = {2022},
  month = jun,
  journal = {Quantum},
  volume = {6},
  pages = {730},
  publisher = {{Verein zur F\"orderung des Open Access Publizierens in den Quantenwissenschaften}},
  doi = {10.22331/q-2022-06-07-730},
  langid = {british}
}

@misc{aliPerformanceStudyVariational2022,
  title = {A {{Performance Study}} of {{Variational Quantum Algorithms}} for {{Solving}} the {{Poisson Equation}} on a {{Quantum Computer}}},
  author = {Ali, Mazen and Kabel, Matthias},
  year = {2022},
  month = nov,
  number = {arXiv:2211.14064},
  eprint = {2211.14064},
  eprinttype = {arxiv},
  primaryclass = {quant-ph},
  publisher = {{arXiv}},
  doi = {10.48550/arXiv.2211.14064},
  archiveprefix = {arXiv}
}

@misc{andersonCommentsUnitaryQubit2022,
  title = {Some {{Comments}} on {{Unitary Qubit Lattice Algorithms}} for {{Classical Problems}}},
  author = {Anderson, Paul and {Finegold-Sachs}, Lillian and Vahala, George and Vahala, Linda and Ram, Abhay K. and Soe, Min and Koukoutsis, Efstratios and Hizandis, Kyriakos},
  year = {2022},
  month = nov,
  number = {arXiv:2211.16661},
  eprint = {2211.16661},
  eprinttype = {arxiv},
  primaryclass = {physics, physics:quant-ph},
  publisher = {{arXiv}},
  archiveprefix = {arXiv},
  langid = {english}
}

@article{anEfficientQuantumAlgorithm2022,
  title = {Efficient Quantum Algorithm for Nonlinear Reaction-Diffusion Equations and Energy Estimation},
  author = {An, Dong and Fang, Di and Jordan, Stephen and Liu, Jin-Peng and Low, Guang Hao and Wang, Jiasu},
  year = {2022},
  publisher = {{arXiv}},
  doi = {10.48550/ARXIV.2205.01141},
  copyright = {arXiv.org perpetual, non-exclusive license}
}

@article{anQuantumacceleratedMultilevelMonte2021,
  title = {Quantum-Accelerated Multilevel {{Monte Carlo}} Methods for Stochastic Differential Equations in Mathematical Finance},
  author = {An, Dong and Linden, Noah and Liu, Jin-Peng and Montanaro, Ashley and Shao, Changpeng and Wang, Jiasu},
  year = {2021},
  month = jun,
  journal = {Quantum},
  volume = {5},
  pages = {481},
  issn = {2521-327X},
  doi = {10.22331/q-2021-06-24-481},
  langid = {english}
}

@article{arrazolaQuantumAlgorithmNonhomogeneous2019,
  title = {Quantum Algorithm for Nonhomogeneous Linear Partial Differential Equations},
  author = {Arrazola, Juan Miguel and Kalajdzievski, Timjan and Weedbrook, Christian and Lloyd, Seth},
  year = {2019},
  month = sep,
  journal = {Physical Review A},
  volume = {100},
  number = {3},
  pages = {032306},
  issn = {2469-9926, 2469-9934},
  doi = {10.1103/PhysRevA.100.032306},
  langid = {english}
}

@article{berryHighorderQuantumAlgorithm2014,
  title = {High-Order Quantum Algorithm for Solving Linear Differential Equations},
  author = {Berry, Dominic W},
  year = {2014},
  month = mar,
  journal = {Journal of Physics A: Mathematical and Theoretical},
  volume = {47},
  number = {10},
  pages = {105301},
  issn = {1751-8113, 1751-8121},
  doi = {10.1088/1751-8113/47/10/105301}
}

@article{berryQuantumAlgorithmLinear2017,
  title = {Quantum {{Algorithm}} for {{Linear Differential Equations}} with {{Exponentially Improved Dependence}} on {{Precision}}},
  author = {Berry, Dominic W. and Childs, Andrew M. and Ostrander, Aaron and Wang, Guoming},
  year = {2017},
  month = dec,
  journal = {Communications in Mathematical Physics},
  volume = {356},
  number = {3},
  pages = {1057--1081},
  issn = {0010-3616, 1432-0916},
  doi = {10.1007/s00220-017-3002-y},
  langid = {english}
}

@misc{borleHowViableQuantum2022,
  title = {How Viable Is Quantum Annealing for Solving Linear Algebra Problems?},
  author = {Borle, Ajinkya and Lomonaco, Samuel J.},
  year = {2022},
  month = jun,
  number = {arXiv:2206.10576},
  eprint = {2206.10576},
  eprinttype = {arxiv},
  primaryclass = {quant-ph},
  publisher = {{arXiv}},
  doi = {10.48550/arXiv.2206.10576},
  archiveprefix = {arXiv}
}

@misc{bravo-prietoVariationalQuantumLinear2020,
  title = {Variational {{Quantum Linear Solver}}},
  author = {{Bravo-Prieto}, Carlos and LaRose, Ryan and Cerezo, M. and Subasi, Yigit and Cincio, Lukasz and Coles, Patrick J.},
  year = {2020},
  month = jun,
  number = {arXiv:1909.05820},
  eprint = {1909.05820},
  eprinttype = {arxiv},
  primaryclass = {quant-ph},
  publisher = {{arXiv}},
  archiveprefix = {arXiv},
  langid = {english}
}

@article{caoQuantumAlgorithmCircuit2013,
  title = {Quantum Algorithm and Circuit Design Solving the {{Poisson}} Equation},
  author = {Cao, Yudong and Papageorgiou, Anargyros and Petras, Iasonas and Traub, Joseph and Kais, Sabre},
  year = {2013},
  month = jan,
  journal = {New Journal of Physics},
  volume = {15},
  number = {1},
  pages = {013021},
  issn = {1367-2630},
  doi = {10.1088/1367-2630/15/1/013021},
  langid = {english}
}

@article{changQuantumAnnealingSystems2019,
  title = {Quantum Annealing for Systems of Polynomial Equations},
  author = {Chang, Chia Cheng and Gambhir, Arjun and Humble, Travis S. and Sota, Shigetoshi},
  year = {2019},
  month = jul,
  journal = {Scientific Reports},
  volume = {9},
  number = {1},
  pages = {10258},
  publisher = {{Nature Publishing Group}},
  issn = {2045-2322},
  doi = {10.1038/s41598-019-46729-0},
  copyright = {2019 The Author(s)},
  langid = {english}
}

@article{childsHighprecisionQuantumAlgorithms2021,
  title = {High-Precision Quantum Algorithms for Partial Differential Equations},
  author = {Childs, Andrew M. and Liu, Jin-Peng and Ostrander, Aaron},
  year = {2021},
  month = nov,
  journal = {Quantum},
  volume = {5},
  eprint = {2002.07868},
  eprinttype = {arxiv},
  primaryclass = {quant-ph},
  pages = {574},
  issn = {2521-327X},
  doi = {10.22331/q-2021-11-10-574},
  archiveprefix = {arXiv},
  langid = {english}
}

@article{childsQuantumAlgorithmSystems2017,
  title = {Quantum {{Algorithm}} for {{Systems}} of {{Linear Equations}} with {{Exponentially Improved Dependence}} on {{Precision}}},
  author = {Childs, Andrew M. and Kothari, Robin and Somma, Rolando D.},
  year = {2017},
  month = jan,
  journal = {SIAM Journal on Computing},
  volume = {46},
  number = {6},
  pages = {1920--1950},
  issn = {0097-5397, 1095-7111},
  doi = {10.1137/16M1087072},
  langid = {english}
}

@article{childsQuantumSpectralMethods2020,
  title = {Quantum {{Spectral Methods}} for {{Differential Equations}}},
  author = {Childs, Andrew M. and Liu, Jin-Peng},
  year = {2020},
  month = apr,
  journal = {Communications in Mathematical Physics},
  volume = {375},
  number = {2},
  pages = {1427--1457},
  issn = {1432-0916},
  doi = {10.1007/s00220-020-03699-z},
  langid = {english}
}

@article{claderPreconditionedQuantumLinear2013,
  title = {Preconditioned {{Quantum Linear System Algorithm}}},
  author = {Clader, B. D. and Jacobs, B. C. and Sprouse, C. R.},
  year = {2013},
  month = jun,
  journal = {Physical Review Letters},
  volume = {110},
  number = {25},
  pages = {250504},
  issn = {0031-9007, 1079-7114},
  doi = {10.1103/PhysRevLett.110.250504},
  langid = {english}
}

@article{costaQuantumAlgorithmSimulating2019,
  title = {Quantum Algorithm for Simulating the Wave Equation},
  author = {Costa, Pedro C. S. and Jordan, Stephen and Ostrander, Aaron},
  year = {2019},
  month = jan,
  journal = {Physical Review A},
  volume = {99},
  number = {1},
  pages = {012323},
  issn = {2469-9926, 2469-9934},
  doi = {10.1103/PhysRevA.99.012323},
  langid = {english}
}

@misc{criadoQadeSolvingDifferential2022,
  title = {Qade: {{Solving Differential Equations}} on {{Quantum Annealers}}},
  shorttitle = {Qade},
  author = {Criado, Juan Carlos and Spannowsky, Michael},
  year = {2022},
  month = apr,
  number = {arXiv:2204.03657},
  eprint = {2204.03657},
  eprinttype = {arxiv},
  primaryclass = {hep-ph, physics:hep-th, physics:quant-ph},
  publisher = {{arXiv}},
  archiveprefix = {arXiv},
  langid = {english}
}

@article{cuiOptimizationNoiseAnalysis2022,
  title = {Optimization and Noise Analysis of the Quantum Algorithm for Solving One-Dimensional {{Poisson}} Equation},
  author = {Cui, Guolong and Wang, Zhimin and Wang, Shengbin and Shi, Shangshang and Shang, Ruimin and Li, Wendong and Wei, Zhiqiang and Gu, Yongjian},
  year = {2022},
  month = may,
  journal = {Quantum Information and Computation},
  volume = {22},
  number = {7\&8},
  pages = {569--593},
  issn = {15337146},
  doi = {10.26421/QIC22.7-8-2}
}

@misc{demirdjianVariationalQuantumSolutions2022,
  title = {Variational {{Quantum Solutions}} to the {{Advection-Diffusion Equation}} for {{Applications}} in {{Fluid Dynamics}}},
  author = {Demirdjian, Reuben and Gunlycke, Daniel and Reynolds, Carolyn A. and Doyle, James D. and Tafur, Sergio},
  year = {2022},
  month = aug,
  number = {arXiv:2208.11780},
  eprint = {2208.11780},
  eprinttype = {arxiv},
  primaryclass = {physics, physics:quant-ph},
  publisher = {{arXiv}},
  doi = {10.48550/arXiv.2208.11780},
  archiveprefix = {arXiv}
}

@article{dodinApplicationsQuantumComputing2021,
  title = {On Applications of Quantum Computing to Plasma Simulations},
  author = {Dodin, I. Y. and Startsev, E. A.},
  year = {2021},
  month = sep,
  journal = {Physics of Plasmas},
  volume = {28},
  number = {9},
  pages = {092101},
  publisher = {{American Institute of Physics}},
  issn = {1070-664X},
  doi = {10.1063/5.0056974}
}

@misc{dodinQuantumComputationNonlinear2021,
  title = {Quantum Computation of Nonlinear Maps},
  author = {Dodin, I. Y. and Startsev, E. A.},
  year = {2021},
  month = may,
  number = {arXiv:2105.07317},
  eprint = {2105.07317},
  eprinttype = {arxiv},
  primaryclass = {physics, physics:quant-ph},
  publisher = {{arXiv}},
  doi = {10.48550/arXiv.2105.07317},
  archiveprefix = {arXiv}
}

@article{engelLinearEmbeddingNonlinear2021,
  title = {Linear Embedding of Nonlinear Dynamical Systems and Prospects for Efficient Quantum Algorithms},
  author = {Engel, Alexander and Smith, Graeme and Parker, Scott E.},
  year = {2021},
  month = jun,
  journal = {Physics of Plasmas},
  volume = {28},
  number = {6},
  pages = {062305},
  publisher = {{American Institute of Physics}},
  issn = {1070-664X},
  doi = {10.1063/5.0040313}
}

@article{engelQuantumAlgorithmVlasov2019,
  title = {Quantum Algorithm for the {{Vlasov}} Equation},
  author = {Engel, Alexander and Smith, Graeme and Parker, Scott E.},
  year = {2019},
  month = dec,
  journal = {Physical Review A},
  volume = {100},
  number = {6},
  pages = {062315},
  publisher = {{American Physical Society}},
  doi = {10.1103/PhysRevA.100.062315}
}

@article{eweVariationalQuantumBasedSimulation2022,
  title = {Variational {{Quantum-Based Simulation}} of {{Waveguide Modes}}},
  author = {Ewe, Wei-Bin and Koh, Dax Enshan and Goh, Siong Thye and Chu, Hong-Son and Png, Ching Eng},
  year = {2022},
  month = may,
  journal = {IEEE Transactions on Microwave Theory and Techniques},
  volume = {70},
  number = {5},
  pages = {2517--2525},
  issn = {1557-9670},
  doi = {10.1109/TMTT.2022.3151510}
}

@misc{fangTimemarchingBasedQuantum2022,
  title = {Time-Marching Based Quantum Solvers for Time-Dependent Linear Differential Equations},
  author = {Fang, Di and Lin, Lin and Tong, Yu},
  year = {2022},
  month = aug,
  number = {arXiv:2208.06941},
  eprint = {2208.06941},
  eprinttype = {arxiv},
  primaryclass = {quant-ph},
  publisher = {{arXiv}},
  doi = {10.48550/arXiv.2208.06941},
  archiveprefix = {arXiv}
}

@article{fontanelaShortCommunicationQuantum2021,
  title = {Short {{Communication}}: {{A Quantum Algorithm}} for {{Linear PDEs Arising}} in {{Finance}}},
  shorttitle = {Short {{Communication}}},
  author = {Fontanela, Filipe and Jacquier, Antoine and Oumgari, Mugad},
  year = {2021},
  month = jan,
  journal = {SIAM Journal on Financial Mathematics},
  volume = {12},
  number = {4},
  pages = {SC98-SC114},
  issn = {1945-497X},
  doi = {10.1137/21M1397878},
  langid = {english}
}

@article{gaitanFindingFlowsNavier2020,
  title = {Finding Flows of a {{Navier}}\textendash{{Stokes}} Fluid through Quantum Computing},
  author = {Gaitan, Frank},
  year = {2020},
  month = jul,
  journal = {npj Quantum Information},
  volume = {6},
  number = {1},
  pages = {1--6},
  publisher = {{Nature Publishing Group}},
  issn = {2056-6387},
  doi = {10.1038/s41534-020-00291-0},
  copyright = {2020 This is a U.S. government work and not under copyright protection in the U.S.; foreign copyright protection may apply},
  langid = {english}
}

@article{garcia-molinaQuantumFourierAnalysis2022,
  title = {Quantum {{Fourier}} Analysis for Multivariate Functions and Applications to a Class of {{Schr}}\textbackslash "odinger-Type Partial Differential Equations},
  author = {{Garc{\'i}a-Molina}, Paula and {Rodr{\'i}guez-Mediavilla}, Javier and {Garc{\'i}a-Ripoll}, Juan Jos{\'e}},
  year = {2022},
  month = jan,
  journal = {Physical Review A},
  volume = {105},
  number = {1},
  pages = {012433},
  publisher = {{American Physical Society}},
  doi = {10.1103/PhysRevA.105.012433}
}

@article{garcia-ripollQuantuminspiredAlgorithmsMultivariate2021,
  title = {Quantum-Inspired Algorithms for Multivariate Analysis: From Interpolation to Partial Differential Equations},
  shorttitle = {Quantum-Inspired Algorithms for Multivariate Analysis},
  author = {{Garc{\'i}a-Ripoll}, Juan Jos{\'e}},
  year = {2021},
  month = apr,
  journal = {Quantum},
  volume = {5},
  eprint = {1909.06619},
  eprinttype = {arxiv},
  pages = {431},
  issn = {2521-327X},
  doi = {10.22331/q-2021-04-15-431},
  archiveprefix = {arXiv},
  langid = {english}
}

@article{harrowQuantumAlgorithmLinear2009,
  title = {Quantum {{Algorithm}} for {{Linear Systems}} of {{Equations}}},
  author = {Harrow, Aram W. and Hassidim, Avinatan and Lloyd, Seth},
  year = {2009},
  month = oct,
  journal = {Physical Review Letters},
  volume = {103},
  number = {15},
  pages = {150502},
  issn = {0031-9007, 1079-7114},
  doi = {10.1103/PhysRevLett.103.150502},
  langid = {english}
}

@article{hatifiQuantumWalkHydrodynamics2019,
  title = {Quantum Walk Hydrodynamics},
  author = {Hatifi, Mohamed and Di Molfetta, Giuseppe and Debbasch, Fabrice and Brachet, Marc},
  year = {2019},
  month = feb,
  journal = {Scientific Reports},
  volume = {9},
  number = {1},
  pages = {2989},
  publisher = {{Nature Publishing Group}},
  issn = {2045-2322},
  doi = {10.1038/s41598-019-40059-x},
  copyright = {2019 The Author(s)},
  langid = {english}
}

@article{huangNeartermQuantumAlgorithms2021,
  title = {Near-Term Quantum Algorithms for Linear Systems of Equations with Regression Loss Functions},
  author = {Huang, Hsin-Yuan and Bharti, Kishor and Rebentrost, Patrick},
  year = {2021},
  month = nov,
  journal = {New Journal of Physics},
  volume = {23},
  number = {11},
  pages = {113021},
  publisher = {{IOP Publishing}},
  issn = {1367-2630},
  doi = {10.1088/1367-2630/ac325f},
  langid = {english}
}

@misc{jinQuantumAlgorithmsComputing2022,
  title = {Quantum Algorithms for Computing Observables of Nonlinear Partial Differential Equations},
  author = {Jin, Shi and Liu, Nana},
  year = {2022},
  month = feb,
  number = {arXiv:2202.07834},
  eprint = {2202.07834},
  eprinttype = {arxiv},
  primaryclass = {physics, physics:quant-ph},
  publisher = {{arXiv}},
  doi = {10.48550/arXiv.2202.07834},
  archiveprefix = {arXiv}
}

@misc{jinQuantumSimulationPartial2022,
  title = {Quantum Simulation of Partial Differential Equations via {{Schrodingerisation}}},
  author = {Jin, Shi and Liu, Nana and Yu, Yue},
  year = {2022},
  month = dec,
  number = {arXiv:2212.13969},
  eprint = {2212.13969},
  eprinttype = {arxiv},
  primaryclass = {quant-ph},
  publisher = {{arXiv}},
  doi = {10.48550/arXiv.2212.13969},
  archiveprefix = {arXiv}
}

@misc{jinQuantumSimulationPartial2022a,
  title = {Quantum Simulation of Partial Differential Equations via {{Schrodingerisation}}: Technical Details},
  shorttitle = {Quantum Simulation of Partial Differential Equations via {{Schrodingerisation}}},
  author = {Jin, Shi and Liu, Nana and Yu, Yue},
  year = {2022},
  month = dec,
  number = {arXiv:2212.14703},
  eprint = {2212.14703},
  eprinttype = {arxiv},
  primaryclass = {quant-ph},
  publisher = {{arXiv}},
  doi = {10.48550/arXiv.2212.14703},
  archiveprefix = {arXiv}
}

@article{jinQuantumSimulationSemiclassical2022,
  title = {Quantum Simulation in the Semi-Classical Regime},
  author = {Jin, Shi and Li, Xiantao and Liu, Nana},
  year = {2022},
  month = jun,
  journal = {Quantum},
  volume = {6},
  pages = {739},
  publisher = {{Verein zur F\"orderung des Open Access Publizierens in den Quantenwissenschaften}},
  doi = {10.22331/q-2022-06-17-739},
  langid = {british}
}

@article{jinTimeComplexityAnalysis2022,
  title = {Time Complexity Analysis of Quantum Algorithms via Linear Representations for Nonlinear Ordinary and Partial Differential Equations},
  author = {Jin, Shi and Liu, Nana and Yu, Yue},
  year = {2022},
  publisher = {{arXiv}},
  doi = {10.48550/ARXIV.2209.08478},
  copyright = {Creative Commons Attribution 4.0 International}
}

@article{jinTimeComplexityAnalysis2022a,
  title = {Time Complexity Analysis of Quantum Difference Methods for Linear High Dimensional and Multiscale Partial Differential Equations},
  author = {Jin, Shi and Liu, Nana and Yu, Yue},
  year = {2022},
  month = dec,
  journal = {Journal of Computational Physics},
  volume = {471},
  pages = {111641},
  issn = {00219991},
  doi = {10.1016/j.jcp.2022.111641},
  langid = {english}
}

@article{josephKoopmanNeumannApproach2020,
  title = {Koopman--von {{Neumann}} Approach to Quantum Simulation of Nonlinear Classical Dynamics},
  author = {Joseph, Ilon},
  year = {2020},
  month = oct,
  journal = {Physical Review Research},
  volume = {2},
  number = {4},
  pages = {043102},
  publisher = {{American Physical Society}},
  doi = {10.1103/PhysRevResearch.2.043102}
}

@misc{josephQuantumComputingFusion2022a,
  title = {Quantum {{Computing}} for {{Fusion Energy Science Applications}}},
  author = {Joseph, I. and Shi, Y. and Porter, M. D. and Castelli, A. R. and Geyko, V. I. and Graziani, F. R. and Libby, S. B. and DuBois, J. L.},
  year = {2022},
  month = dec,
  number = {arXiv:2212.05054},
  eprint = {2212.05054},
  eprinttype = {arxiv},
  primaryclass = {math-ph, physics:physics, physics:quant-ph},
  publisher = {{arXiv}},
  doi = {10.48550/arXiv.2212.05054},
  archiveprefix = {arXiv}
}

@article{joubert-doriolVariationalApproachLinearly2022,
  title = {A Variational Approach for Linearly Dependent Moving Bases in Quantum Dynamics: Application to {{Gaussian}} Functions},
  shorttitle = {A Variational Approach for Linearly Dependent Moving Bases in Quantum Dynamics},
  author = {{Joubert-Doriol}, Lo{\"i}c},
  year = {2022},
  month = may,
  journal = {arXiv:2205.02358 [physics, physics:quant-ph]},
  eprint = {2205.02358},
  eprinttype = {arxiv},
  primaryclass = {physics, physics:quant-ph},
  archiveprefix = {arXiv}
}

@misc{koukoutsisDysonMapsUnitary2022,
  title = {Dyson {{Maps}} and {{Unitary Evolution}} for {{Maxwell Equations}} in {{Tensor Dielectric Media}}},
  author = {Koukoutsis, Efstratios and Hizanidis, Kyriakos and Ram, Abhay K. and Vahala, George},
  year = {2022},
  month = sep,
  number = {arXiv:2209.08523},
  eprint = {2209.08523},
  eprinttype = {arxiv},
  primaryclass = {physics, physics:quant-ph},
  publisher = {{arXiv}},
  doi = {10.48550/arXiv.2209.08523},
  archiveprefix = {arXiv}
}

@article{kroviImprovedQuantumAlgorithms2022,
  title = {Improved Quantum Algorithms for Linear and Nonlinear Differential Equations},
  author = {Krovi, Hari},
  year = {2022},
  month = feb,
  journal = {arXiv:2202.01054 [physics, physics:quant-ph]},
  eprint = {2202.01054},
  eprinttype = {arxiv},
  primaryclass = {physics, physics:quant-ph},
  archiveprefix = {arXiv}
}

@article{kuboVariationalQuantumSimulations2020,
  title = {Variational Quantum Simulations of Stochastic Differential Equations},
  author = {Kubo, Kenji and Nakagawa, Yuya O. and Endo, Suguru and Nagayama, Shota},
  year = {2020},
  month = dec,
  journal = {arXiv:2012.04429 [quant-ph]},
  eprint = {2012.04429},
  eprinttype = {arxiv},
  primaryclass = {quant-ph},
  archiveprefix = {arXiv}
}

@article{kyriienkoSolvingNonlinearDifferential2021,
  title = {Solving Nonlinear Differential Equations with Differentiable Quantum Circuits},
  author = {Kyriienko, Oleksandr and Paine, Annie E. and Elfving, Vincent E.},
  year = {2021},
  month = may,
  journal = {Physical Review A},
  volume = {103},
  number = {5},
  pages = {052416},
  publisher = {{American Physical Society}},
  doi = {10.1103/PhysRevA.103.052416}
}

@misc{lapworthHybridQuantumClassicalCFD2022,
  title = {A {{Hybrid Quantum-Classical CFD Methodology}} with {{Benchmark HHL Solutions}}},
  author = {Lapworth, Leigh},
  year = {2022},
  month = jun,
  number = {arXiv:2206.00419},
  eprint = {2206.00419},
  eprinttype = {arxiv},
  primaryclass = {quant-ph},
  publisher = {{arXiv}},
  doi = {10.48550/arXiv.2206.00419},
  archiveprefix = {arXiv}
}

@article{leongVariationalQuantumEvolution2022,
  title = {Variational {{Quantum Evolution Equation Solver}}},
  author = {Leong, Fong Yew and Ewe, Wei-Bin and Koh, Dax Enshan},
  year = {2022},
  month = dec,
  journal = {Scientific Reports},
  volume = {12},
  number = {1},
  eprint = {2204.02912},
  eprinttype = {arxiv},
  primaryclass = {physics, physics:quant-ph},
  pages = {10817},
  issn = {2045-2322},
  doi = {10.1038/s41598-022-14906-3},
  archiveprefix = {arXiv},
  langid = {english}
}

@misc{leytonQuantumAlgorithmSolve2008,
  title = {A Quantum Algorithm to Solve Nonlinear Differential Equations},
  author = {Leyton, Sarah K. and Osborne, Tobias J.},
  year = {2008},
  month = dec,
  number = {arXiv:0812.4423},
  eprint = {0812.4423},
  eprinttype = {arxiv},
  primaryclass = {quant-ph},
  publisher = {{arXiv}},
  doi = {10.48550/arXiv.0812.4423},
  archiveprefix = {arXiv}
}

@article{lindenQuantumVsClassical2022,
  title = {Quantum vs. {{Classical Algorithms}} for {{Solving}} the {{Heat Equation}}},
  author = {Linden, Noah and Montanaro, Ashley and Shao, Changpeng},
  year = {2022},
  month = aug,
  journal = {Communications in Mathematical Physics},
  issn = {1432-0916},
  doi = {10.1007/s00220-022-04442-6},
  langid = {english}
}

@misc{linKoopmanNeumannMechanics2022,
  title = {Koopman von {{Neumann}} Mechanics and the {{Koopman}} Representation: {{A}} Perspective on Solving Nonlinear Dynamical Systems with Quantum Computers},
  shorttitle = {Koopman von {{Neumann}} Mechanics and the {{Koopman}} Representation},
  author = {Lin, Yen Ting and Lowrie, Robert B. and Aslangil, Denis and Suba{\c s}{\i}, Yi{\u g}it and Sornborger, Andrew T.},
  year = {2022},
  month = feb,
  number = {arXiv:2202.02188},
  eprint = {2202.02188},
  eprinttype = {arxiv},
  primaryclass = {quant-ph},
  publisher = {{arXiv}},
  doi = {10.48550/arXiv.2202.02188},
  archiveprefix = {arXiv}
}

@article{liuApplicationVariationalHybrid2022,
  title = {Application of a Variational Hybrid Quantum-Classical Algorithm to Heat Conduction Equation and Analysis of Time Complexity},
  author = {Liu, Y. Y. and Chen, Z. and Shu, C. and Chew, S. C. and Khoo, B. C. and Zhao, X. and Cui, Y. D.},
  year = {2022},
  month = nov,
  journal = {Physics of Fluids},
  volume = {34},
  number = {11},
  pages = {117121},
  publisher = {{American Institute of Physics}},
  issn = {1070-6631},
  doi = {10.1063/5.0121778}
}

@article{liuEfficientQuantumAlgorithm2021,
  title = {Efficient Quantum Algorithm for Dissipative Nonlinear Differential Equations},
  author = {Liu, Jin-Peng and Kolden, Herman {\O}ie and Krovi, Hari K. and Loureiro, Nuno F. and Trivisa, Konstantina and Childs, Andrew M.},
  year = {2021},
  month = aug,
  journal = {Proceedings of the National Academy of Sciences},
  volume = {118},
  number = {35},
  pages = {e2026805118},
  publisher = {{Proceedings of the National Academy of Sciences}},
  doi = {10.1073/pnas.2026805118}
}

@article{liuVariationalQuantumAlgorithm2021,
  title = {Variational Quantum Algorithm for the {{Poisson}} Equation},
  author = {Liu, Hai-Ling and Wu, Yu-Sen and Wan, Lin-Chun and Pan, Shi-Jie and Qin, Su-Juan and Gao, Fei and Wen, Qiao-Yan},
  year = {2021},
  month = aug,
  journal = {Physical Review A},
  volume = {104},
  number = {2},
  pages = {022418},
  issn = {2469-9926, 2469-9934},
  doi = {10.1103/PhysRevA.104.022418},
  langid = {english}
}

@article{ljubomirQuantumAlgorithmNavier2022,
  title = {Quantum Algorithm for the {{Navier}}\textendash{{Stokes}} Equations by Using the Streamfunction-Vorticity Formulation and the Lattice {{Boltzmann}} Method},
  author = {Ljubomir, Budinski},
  year = {2022},
  month = mar,
  journal = {International Journal of Quantum Information},
  volume = {20},
  number = {02},
  pages = {2150039},
  publisher = {{World Scientific Publishing Co.}},
  issn = {0219-7499},
  doi = {10.1142/S0219749921500398}
}

@misc{lloydQuantumAlgorithmNonlinear2020,
  title = {Quantum Algorithm for Nonlinear Differential Equations},
  author = {Lloyd, Seth and De Palma, Giacomo and Gokler, Can and Kiani, Bobak and Liu, Zi-Wen and Marvian, Milad and Tennie, Felix and Palmer, Tim},
  year = {2020},
  month = dec,
  number = {arXiv:2011.06571},
  eprint = {2011.06571},
  eprinttype = {arxiv},
  primaryclass = {nlin, physics:quant-ph},
  publisher = {{arXiv}},
  doi = {10.48550/arXiv.2011.06571},
  archiveprefix = {arXiv}
}

@article{lubaschVariationalQuantumAlgorithms2020,
  title = {Variational Quantum Algorithms for Nonlinear Problems},
  author = {Lubasch, Michael and Joo, Jaewoo and Moinier, Pierre and Kiffner, Martin and Jaksch, Dieter},
  year = {2020},
  month = jan,
  journal = {Physical Review A},
  volume = {101},
  number = {1},
  pages = {010301},
  issn = {2469-9926, 2469-9934},
  doi = {10.1103/PhysRevA.101.010301},
  langid = {english}
}

@article{miyamotoPricingMultiAssetDerivatives2022,
  title = {Pricing {{Multi-Asset Derivatives}} by {{Finite-Difference Method}} on a {{Quantum Computer}}},
  author = {Miyamoto, Koichi and Kubo, Kenji},
  year = {2022},
  journal = {IEEE Transactions on Quantum Engineering},
  volume = {3},
  pages = {1--25},
  issn = {2689-1808},
  doi = {10.1109/TQE.2021.3128643}
}

@article{moczCosmologicalSimulationsDark2021,
  title = {Toward {{Cosmological Simulations}} of {{Dark Matter}} on {{Quantum Computers}}},
  author = {Mocz, Philip and Szasz, Aaron},
  year = {2021},
  month = mar,
  journal = {The Astrophysical Journal},
  volume = {910},
  number = {1},
  pages = {29},
  publisher = {{The American Astronomical Society}},
  issn = {0004-637X},
  doi = {10.3847/1538-4357/abe6ac},
  langid = {english}
}

@article{montanaroQuantumAlgorithmsFinite2016,
  title = {Quantum Algorithms and the Finite Element Method},
  author = {Montanaro, Ashley and Pallister, Sam},
  year = {2016},
  month = mar,
  journal = {Physical Review A},
  volume = {93},
  number = {3},
  pages = {032324},
  publisher = {{American Physical Society}},
  doi = {10.1103/PhysRevA.93.032324}
}

@article{novikauQuantumSignalProcessing2022,
  title = {Quantum Signal Processing for Simulating Cold Plasma Waves},
  author = {Novikau, I. and Startsev, E. A. and Dodin, I. Y.},
  year = {2022},
  month = jun,
  journal = {Physical Review A},
  volume = {105},
  number = {6},
  pages = {062444},
  publisher = {{American Physical Society}},
  doi = {10.1103/PhysRevA.105.062444}
}

@misc{novikauSimulationLinearNonHermitian2022,
  title = {Simulation of Linear Non-{{Hermitian}} Boundary-Value Problems with Quantum Singular Value Transformation},
  author = {Novikau, I. and Dodin, I. Y. and Startsev, E. A.},
  year = {2022},
  month = dec,
  number = {arXiv:2212.09113},
  eprint = {2212.09113},
  eprinttype = {arxiv},
  primaryclass = {physics, physics:quant-ph},
  publisher = {{arXiv}},
  doi = {10.48550/arXiv.2212.09113},
  archiveprefix = {arXiv}
}

@article{oganesovBenchmarkingDiracgeneratedUnitary2016,
  title = {Benchmarking the {{Dirac-generated}} Unitary Lattice Qubit Collision-Stream Algorithm for {{1D}} Vector {{Manakov}} Soliton Collisions},
  author = {Oganesov, Armen and Vahala, George and Vahala, Linda and Yepez, Jeffrey and Soe, Min},
  year = {2016},
  month = jul,
  journal = {Computers \& Mathematics with Applications},
  volume = {72},
  number = {2},
  pages = {386--393},
  issn = {08981221},
  doi = {10.1016/j.camwa.2015.06.001},
  langid = {english}
}

@article{oganesovEffectFourierTransform2018,
  title = {Effect of {{Fourier}} Transform on the Streaming in Quantum Lattice Gas Algorithms},
  author = {Oganesov, Armen and Vahala, George and Vahala, Linda and Soe, Min},
  year = {2018},
  month = apr,
  journal = {Radiation Effects and Defects in Solids},
  volume = {173},
  number = {3-4},
  pages = {169--174},
  publisher = {{Taylor \& Francis}},
  issn = {1042-0150},
  doi = {10.1080/10420150.2018.1462364}
}

@article{oganesovImaginaryTimeIntegration2016,
  title = {Imaginary Time Integration Method Using a Quantum Lattice Gas Approach},
  author = {Oganesov, Armen and Flint, Christopher and Vahala, George and Vahala, Linda and Yepez, Jeffrey and Soe, Min},
  year = {2016},
  month = feb,
  journal = {Radiation Effects and Defects in Solids},
  volume = {171},
  number = {1-2},
  pages = {96--102},
  publisher = {{Taylor \& Francis}},
  issn = {1042-0150},
  doi = {10.1080/10420150.2015.1137916}
}

@article{oganesovUnitaryQuantumLattice2015,
  title = {Unitary Quantum Lattice Gas Algorithm Generated from the {{Dirac}} Collision Operator for {{1D}} Soliton\textendash Soliton Collisions},
  author = {Oganesov, Armen and Vahala, George and Vahala, Linda and Yepez, Jeffrey and Soe, Min and Zhang, Bo},
  year = {2015},
  month = jan,
  journal = {Radiation Effects and Defects in Solids},
  volume = {170},
  number = {1},
  pages = {55--64},
  publisher = {{Taylor \& Francis}},
  issn = {1042-0150},
  doi = {10.1080/10420150.2014.988625}
}

@article{omalleyNeartermQuantumAlgorithm2022,
  title = {A Near-Term Quantum Algorithm for Solving Linear Systems of Equations Based on the {{Woodbury}} Identity},
  author = {O'Malley, Daniel and Subasi, Yigit and Golden, John and Lowrie, Robert and Eidenbenz, Stephan},
  year = {2022},
  month = may,
  journal = {arXiv:2205.00645 [quant-ph]},
  eprint = {2205.00645},
  eprinttype = {arxiv},
  primaryclass = {quant-ph},
  archiveprefix = {arXiv}
}

@article{ozSolvingBurgersEquation2021,
  title = {Solving {{Burgers}}' Equation with Quantum Computing},
  author = {Oz, Furkan and Vuppala, Rohit K. S. S. and Kara, Kursat and Gaitan, Frank},
  year = {2021},
  month = dec,
  journal = {Quantum Information Processing},
  volume = {21},
  number = {1},
  pages = {30},
  issn = {1573-1332},
  doi = {10.1007/s11128-021-03391-8},
  langid = {english}
}

@misc{patelQuantumInspiredTensorNeural2022,
  title = {Quantum-{{Inspired Tensor Neural Networks}} for {{Partial Differential Equations}}},
  author = {Patel, Raj and Hsing, Chia-Wei and Sahin, Serkan and Jahromi, Saeed S. and Palmer, Samuel and Sharma, Shivam and Michel, Christophe and Porte, Vincent and Abid, Mustafa and Aubert, Stephane and Castellani, Pierre and Lee, Chi-Guhn and Mugel, Samuel and Orus, Roman},
  year = {2022},
  month = aug,
  number = {arXiv:2208.02235},
  eprint = {2208.02235},
  eprinttype = {arxiv},
  primaryclass = {cond-mat, physics:physics, physics:quant-ph},
  publisher = {{arXiv}},
  doi = {10.48550/arXiv.2208.02235},
  archiveprefix = {arXiv}
}

@article{ramReflectionTransmissionElectromagnetic2021,
  title = {Reflection and Transmission of Electromagnetic Pulses at a Planar Dielectric Interface: {{Theory}} and Quantum Lattice Simulations},
  shorttitle = {Reflection and Transmission of Electromagnetic Pulses at a Planar Dielectric Interface},
  author = {Ram, Abhay K. and Vahala, George and Vahala, Linda and Soe, Min},
  year = {2021},
  month = oct,
  journal = {AIP Advances},
  volume = {11},
  number = {10},
  pages = {105116},
  publisher = {{American Institute of Physics}},
  doi = {10.1063/5.0067204}
}

@article{ricardoAlternativesNonhomogeneousPartial2022,
  title = {Alternatives to a Nonhomogeneous Partial Differential Equation Quantum Algorithm},
  author = {Ricardo, Alexandre C. and Fernandes, Gabriel P. L. M. and Duzzioni, Eduardo I. and Campo, Vivaldo L. and {Villas-Boas}, Celso J.},
  year = {2022},
  month = nov,
  journal = {Physical Review A},
  volume = {106},
  number = {5},
  pages = {052431},
  publisher = {{American Physical Society}},
  doi = {10.1103/PhysRevA.106.052431}
}

@misc{sahaAdvancingAlgorithmScale2022,
  title = {Advancing {{Algorithm}} to {{Scale}} and {{Accurately Solve Quantum Poisson Equation}} on {{Near-term Quantum Hardware}}},
  author = {Saha, Kamal K. and Robson, Walter and Howington, Connor and Suh, In-Saeng and Wang, Zhimin and Nabrzyski, Jaroslaw},
  year = {2022},
  month = oct,
  number = {arXiv:2210.16668},
  eprint = {2210.16668},
  eprinttype = {arxiv},
  primaryclass = {quant-ph},
  publisher = {{arXiv}},
  doi = {10.48550/arXiv.2210.16668},
  archiveprefix = {arXiv}
}

@article{sarsengeldinHybridClassicalQuantumFramework2022,
  title = {A {{Hybrid Classical-Quantum}} Framework for Solving {{Free Boundary Value Problems}} and {{Applications}} in {{Modeling Electric Contact Phenomena}}},
  author = {Sarsengeldin, Merey M.},
  year = {2022},
  month = may,
  journal = {arXiv:2205.02230 [quant-ph]},
  eprint = {2205.02230},
  eprinttype = {arxiv},
  primaryclass = {quant-ph},
  archiveprefix = {arXiv}
}

@article{satoVariationalQuantumAlgorithm2021,
  title = {Variational Quantum Algorithm Based on the Minimum Potential Energy for Solving the {{Poisson}} Equation},
  author = {Sato, Yuki and Kondo, Ruho and Koide, Satoshi and Takamatsu, Hideki and Imoto, Nobuyuki},
  year = {2021},
  month = nov,
  journal = {Physical Review A},
  volume = {104},
  number = {5},
  pages = {052409},
  issn = {2469-9926, 2469-9934},
  doi = {10.1103/PhysRevA.104.052409},
  langid = {english}
}

@article{shaoFasterQuantuminspiredAlgorithms2021,
  title = {Faster Quantum-Inspired Algorithms for Solving Linear Systems},
  author = {Shao, Changpeng and Montanaro, Ashley},
  year = {2021},
  month = mar,
  journal = {arXiv:2103.10309 [quant-ph]},
  eprint = {2103.10309},
  eprinttype = {arxiv},
  primaryclass = {quant-ph},
  archiveprefix = {arXiv}
}

@article{shiSimulatingNonnativeCubic2021,
  title = {Simulating Non-Native Cubic Interactions on Noisy Quantum Machines},
  author = {Shi, Yuan and Castelli, Alessandro R. and Wu, Xian and Joseph, Ilon and Geyko, Vasily and Graziani, Frank R. and Libby, Stephen B. and Parker, Jeffrey B. and Rosen, Yaniv J. and Martinez, Luis A. and DuBois, Jonathan L.},
  year = {2021},
  month = jun,
  journal = {Physical Review A},
  volume = {103},
  number = {6},
  pages = {062608},
  publisher = {{American Physical Society}},
  doi = {10.1103/PhysRevA.103.062608}
}

@article{vahalaQuantumLatticeGas2003,
  title = {Quantum Lattice Gas Representation of Some Classical Solitons},
  author = {Vahala, George and Yepez, Jeffrey and Vahala, Linda},
  year = {2003},
  month = apr,
  journal = {Physics Letters A},
  volume = {310},
  number = {2},
  pages = {187--196},
  issn = {0375-9601},
  doi = {10.1016/S0375-9601(03)00334-7},
  langid = {english}
}

\end{document}